\documentclass[fp,twocolumn]{jpsj3}
\usepackage{txfonts}
\usepackage[version=3]{mhchem}
\usepackage{epstopdf}
\usepackage{color}
\usepackage{epsf}
\usepackage{graphicx}
\usepackage{dcolumn}
\usepackage{bm}
\usepackage{soul}
\usepackage{color}
\usepackage[version=3]{mhchem}
\usepackage{lipsum}
\usepackage[outercaption]{sidecap}
\usepackage{floatrow}
\title{Thermal-incuded stress of plasmonic magnetic nanocomposites}

\author{Anh D. Phan$^{1,2}$\thanks{adphan35@gmail.com}, Nghia C. Do$^3$, and Do T. Nga$^1$}
\inst{$^1$Institute of Physics, Vietnam Academy of Science and Technology, 10 Dao Tan, Ba Dinh, Hanoi, Vietnam\\
$^2$Department of Physics, University of Illinois, 1110 West Green St, Urbana, Illinois 61801, USA\\
$^3$Hanoi Pedagogical University 2, Nguyen Van Linh Street, Vinh Phuc, Vietnam}
\abst{We present theoretical calculations to interpret optical and mechanical properties of Ag@\ce{Fe_3O_4} nanoflowers. The microstructures and nature of optical peaks of nanoflowers are determined by means of the Mie theory associated with effective dielectric approximation and the experimental absorption spectrum. Under laser illumination, the thermal strain fields inside and outside the structure due to the absorbed optical energy are studied using continuum mechanics approach. Our findings provide simple but comprehensive description of the elastic behaviors of previous experiments.}

\begin{document}
\maketitle

\section{Introduction}

In recent years, peculiar properties of composite nanostructures have become of great interest because they have various technological and biomedical applications \cite{1,5,6,8,10,12,13}. The development of nanoscience allows the synthesis of a variety of hybrid systems with desired sizes and structures enabling the designing and creation of multifunctional and complex systems. Magnetite nanoparticles possess novel characteristics such as strong sensitivity to magnetic fields \cite{10}, high biocompatibility and relatively low toxicity in human body, along with capability of removing heavy metal ions \cite{11}. While the localized surface plasmon resonances of silver nanoparticles greatly enhance electric field around the nanostructures and induce other fascinating features in visible range. Silver nanoparticles are also known as effectively antibacterial agents but toxic metal \cite{14}. Local heating of Ag@\ce{Fe_3O_4} nanocomposites using both photothermal and magnetic hyperthermia effects have been demonstrated to remarkably increase a rapid temperature rise compared to the single method \cite{10}, and can be used for cancer treatment. Moreover, the localized heat generates spatiotemporally heterogeneous temperature distribution and can improve ultrasonic imaging \cite{18,19}. Consequently, combining Ag and \ce{Fe_3O_4} has considerably enlarged desirable synergistic and complementary effects. In this paper, we theoretically investigate optical and elastic properties of Ag@\ce{Fe_3O_4} core-shell nanoflowers (Fig.\ref{fig:0}), which has been experimentally synthesized in Ref.\cite{10} in visible range. 

\section{Optical absorption properties of Ag@\ce{Fe_3O_4} nanoflowers}
The Mie theory has been widely used to study the optical response of a spherical nanoparticle with arbitrary size and shows a good agreement with experimental results \cite{1}. When a radius of the nanoparticle is much smaller than considered wavelength, the Mie coefficients can be simplified by the quasi-static approximation. Using the exact solution of the Mie theory, thus, is needed to calculate accurately the absorption cross section of the composite nanoparticle. Exact expressions of the Mie scattering theory calculating the extinction, scattering, and absorption cross section of isotropically coated spherical nanoparticles are given by \cite{7}

\begin{eqnarray}
Q_{ext} &=& -\frac{2\pi}{k_m^2}\sum_{n=1}^{\infty}(2n+1)\ce{Re}\left(a_n + b_n \right), \nonumber\\
Q_{scat} &=& \frac{2\pi}{k_m^2}\sum_{n=1}^{\infty}(2n+1)\left(\left | a_n \right|^2  + \left | b_n \right |^2 \right),\nonumber\\
Q_{abs} &=& Q_{ext}  - Q_{scat},
\label{eq:1}
\end{eqnarray}

where 
\begin{eqnarray}
a_n=-\frac{U_n^{TM}}{U_n^{TM}+iV_n^{TM}} ,\quad b_n=-\frac{U_n^{TE}}{U_n^{TE}+iV_n^{TE}}, \nonumber\\
U_n^{TM} = \begin{vmatrix} j_n(k_cR_c) & j_n(k_sR_c) & y_n(k_sR_c) & 0 \\ \cfrac{\Psi_n^{'}(k_cR_c)}{\varepsilon_c} & \cfrac{\Psi_n^{'}(k_sR_c)}{\varepsilon_s} & \cfrac{\Phi_n^{'}(k_sR_c)}{\varepsilon_s}& 0 \\ 
0 & j_n(k_sR_s) & y_n(k_sR_s) & j_n(k_mR_s) \\
0 & \cfrac{\Psi_n^{'}(k_sR_s)}{\varepsilon_c} & \cfrac{\Phi_n^{'}(k_sR_s)}{\varepsilon_s} & \cfrac{\Psi_n^{'}(k_mR_s)}{\varepsilon_m}
\end{vmatrix}, \nonumber\\
V_n^{TM} = \begin{vmatrix} j_n(k_cR_c) & j_n(k_sR_c) & y_n(k_sR_c) & 0 \\ \cfrac{\Psi_n^{'}(k_cR_c)}{\varepsilon_c} & \cfrac{\Psi_n^{'}(k_sR_c)}{\varepsilon_s} & \cfrac{\Phi_n^{'}(k_sR_c)}{\varepsilon_s}& 0 \\ 
0 & j_n(k_sR_s) & y_n(k_sR_s) & y_n(k_mR_s) \\
0 & \cfrac{\Psi_n^{'}(k_sR_s)}{\varepsilon_c} & \cfrac{\Phi_n^{'}(k_sR_s)}{\varepsilon_s} & \cfrac{\Phi_n^{'}(k_mR_s)}{\varepsilon_m}
\end{vmatrix},\nonumber\\
\label{eq:2}
\end{eqnarray}
where $V_n$ and $U_n$ are determinants, $j_n(x)$ is the spherical Bessel function of the first kind, $y_n(x)$ is the spherical Neumann function, $\Psi(x)=xj_n(x)$ and $\xi_n(x) = xy_n(x)$ are the Riccati–Bessel functions. $\varepsilon_c$, $\varepsilon_s$ and $\varepsilon_m$ are the dielectric functions of core, shell and surrounding medium of the nanocomposite, respectively.  $U_n^{TE}$ and $V_n^{TE}$ can be obtained by replacing the dielectric function by the permeability in $U_n^{TM}$ and $V_n^{TM}$ of Eq. (\ref{eq:2}). The wavenumber is $k_i=2\pi\sqrt{\varepsilon_i}/\lambda$ with $i=s, c$, and $m$. $\lambda$ is the wavelength of incident light in vacuum. 
Since all materials have a trivial permeability $\mu=1$ within the visible range, effects of magnetic properties of \ce{Fe_3O_4} on optical spectrum is ignorable. 

\begin{figure}[htp]
\includegraphics[width=4.5cm]{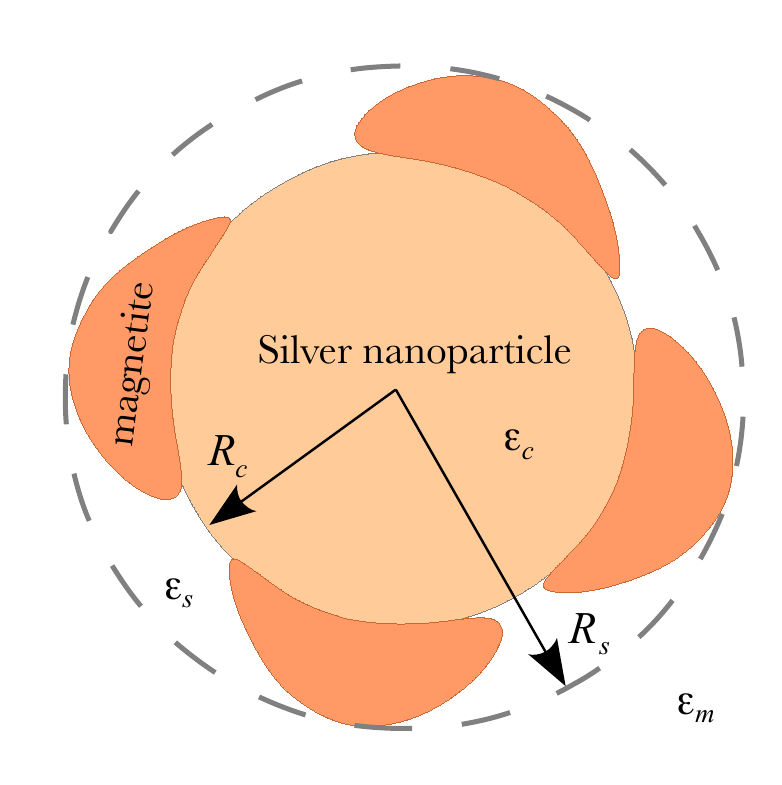}
\caption{\label{fig:0}(Color online) Schematic illustration of Ag@\ce{Fe_3O_4} nanoflowers.}
\end{figure}

In the nanoflowers synthesized in Ref.\cite{10}, the iron oxide petals are grown on silver surface, thus the shell includes \ce{Fe_3O_4} and medium (seen in Fig.\ref{fig:0}). The effective dielectric function of the shell layer can be modeled by the Lichtenecker model \cite{33,34,35} $\varepsilon_s(\omega)=f\varepsilon_{\ce{Fe_3O_4}}+(1-f)\varepsilon_m$, where $f$ is the fraction of magnetite comprising the shell. While $\varepsilon_c(\omega)=\varepsilon_{Ag}(\omega)$. $R_{c} = 24$ nm is a radius of the silver core and $R_{s} = 60$ nm is an outer radius of the nanocomposite.

The dielectric function of Ag is obtained by fitting experimental data in \cite{2} with the Lorentz-Drude model
\begin{eqnarray}
\varepsilon_c(\omega)= 2-\frac{f_0\omega_p^2}{\omega^2 + i\omega(\gamma_0+\frac{Av_F}{R_{c}})}+\frac{f_1\omega_p^2}{\omega_j^2 -i\omega\gamma_j-\omega^2}, 
\label{3}
\end{eqnarray}
where $v_F$ is the Fermi velocity of silver, $A = 0.6$ is a parameter characterizing for the size effect on the plasmonic property, $f_0=1.04$ and $f_1=1.13$ are the oscillator strengths, $\omega_p=9.01$ eV is the plasma frequency of Ag,  $\omega_1=5.98$ eV is a Lorentz frequency corresponding to first resonance, damping parameters are $\gamma_0=0.0165$ eV and $\gamma_1=0.43$ eV. The dielectric function of $\ce{Fe_3O_4}$ and its parameters are taken from Ref.\cite{3}.

\begin{figure}[htp]
\includegraphics[width=8cm]{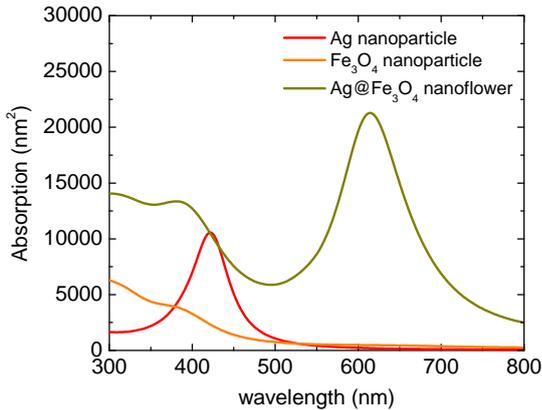}
\caption{\label{fig:1}(Color online) Absorption cross sections of  Ag, \ce{Fe_3O_4} and Ag@\ce{Fe_3O_4} nanoparticle in water ($\varepsilon_m = 1.77$) calculated by general Mie theory.}
\end{figure}

Figure \ref{fig:1} shows theoretical calculations using the Mie theory for the absorption spectrum of Ag@\ce{Fe_3O_4} nanoflowers, pure spherical Ag and \ce{Fe_3O_4} nanoparticles dispersed in water. The resonance peak wavelength of silver nanoparticles with a radius of 24 nm is around 415 nm. The surface plasmon resonance of silver and the electronic transition of magnetite at 400 nm \cite{28,29,4} are mainly responsible for the first band of Ag@\ce{Fe_3O_4} nanoflowers. While the 2.2 eV ($\sim$ 565 nm) band gap of magnetite nanostructure significantly reduces the effects of lower energy excitations on the absorption spectrum \cite{32}, it results in an absence of an optical maximum of \ce{Fe_3O_4} nanoparticles experimentally observed and theoretically calculated in visible range \cite{28,29}. When magnetite is coated on the surface of silver nanoparticles, the surface defects can narrow the band gap of magnetite petal \cite{32} and lead to the occurence of the second peak in the absorption spectrum of Ag@\ce{Fe_3O_4} nanocomposites. Within the framework of the Mie theory, the dielectric function of $Fe_3O_4$ is integrated in $Q_{abs}$ via spatial averaging. Consequently, the second band can be interpreted as geometric effects and the interfacial interaction plays minor role in the absorption spectrum. To have the best agreement between our theoretical calculations and previous experiment \cite{10}, an adjustable parameter $f$ is altered to achieve the second peak at 620 nm (close to 615 nm in Ref.\cite{10}) and the 620 nm absorbance higher than the 400 nm resonance by a factor of 1.5. Based on these criteria, magnetite is found to cover 40 $\%$ volume of the shell ($f \approx 0.4$).

\begin{figure}[htp]
\includegraphics[width=8cm]{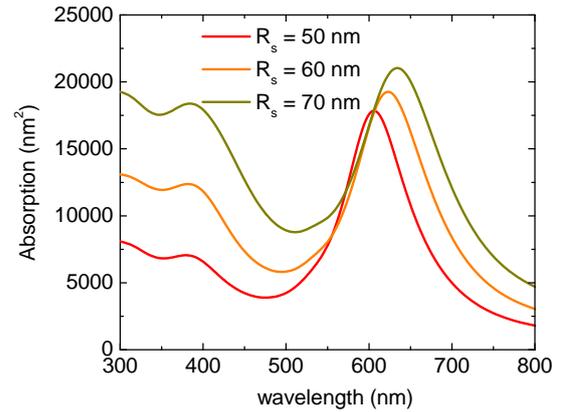}
\caption{\label{fig:2}(Color online) Absorption cross sections of  Ag@\ce{Fe_3O_4} nanoflowers calculated by general Mie theory in silica ($\varepsilon_m = 2.25$) with different shell sizes.}
\end{figure}

Figure \ref{fig:2} presents how the absorption spectra of nanoflowers in silica are altered when the magnetite size varies ($R_c = 24$ nm). When replacing water medium with silica, the small dielectric constant difference leads to a red-shift in the second plasmonic resonance toward the near-infrared region and a relatively small blue-shift in the first absorption peak. An increase in the thickness of the magnetite layer remarkably increases the absorption intensity near 400 nm wavelength due to coupling between the electronic transitions of magnetite and the plasmon band of silver. Alternatively, a crude model to explain absorption enhancement can be made by considering the nanoflowers as effective one-component spherical nanoparticles. In spite of the fact that the dielectric function of the effective nanoparticles is implicitly dependent on $R_s$, it is well known in the Mie theory that $Q_{abs} \sim 4\pi R_s^3$ \cite{1}. Thus, if we suppose that influence of $R_s$ on the effective dielectric function is small, the absorption cross section grows with the shell thickness.

\section{Thermal strain of  Ag@\ce{Fe_3O_4} nanoflowers}
As the Ag@\ce{Fe_3O_4} suspension is exposed to 400-nm laser light, nanoflowers absorb optical energy at the plasmonic resonant wavelength and effectively generate heat. For the steady state, the distribution of the radially symmetric temperature rise outside a nanoflower ($r\ge R_s$) is given \cite{15}
\begin{eqnarray}
T(r) = \frac{Q_{abs}I_0}{4\pi \kappa_m}\frac{1}{r} = T_s\frac{R_s}{r},
\label{eq:5}
\end{eqnarray}
where $\kappa_m=0.6$ $Wm^{-1}K^{-1}$ is the thermal conductivity of water, $T_s$ is the temperature rise on the surface of the nanoflower, and $I_0$ is the intensity of the exposing laser light. The temperature profile inside nanoflowers can be obtained using the heat diffusion conditions at interfaces
\begin{eqnarray} 
\left\{ \begin{array}{rcl}
-T_s \left[ \cfrac{\kappa_m r^2}{2\kappa_cR_s^2}-\cfrac{2\kappa_s+\kappa_m}{2\kappa_s}+\cfrac{R_c^2\kappa_m(\kappa_c-\kappa_s)}{2R_s^2\kappa_s\kappa_c}\right] & \mbox{in core}&  \\ 
T_s\left[-\cfrac{\kappa_m r^2}{2\kappa_sR_s^2} +\cfrac{2\kappa_s+\kappa_m}{2\kappa_s}\right] & \mbox{in shell} &  
\end{array}\right.
\label{inside}
\end{eqnarray}
where $\kappa_s = f\kappa_{Fe_3O_4}+(1-f)\kappa_m \approx 2.76$ $Wm^{-1}K^{-1}$ and $\kappa_c=430$ $Wm^{-1}K^{-1}$ are the thermal conductivity of shell and core, respectively, and $\kappa_{Fe_3O_4}\approx 6$ $Wm^{-1}K^{-1}$ is the thermal conductivity of pure magnetite \cite{30,31}. Since $\kappa_c \gg \kappa_m$, the first term of the core temperature in Eq.(\ref{inside}) can be ignored, so $T(r)$ in the core and shell are relatively different but approximately constant.

Since nanoflowers are modeled to be a roughly spherical shape and are subject to thermal gradients, we assume the deformation induced by thermoelastic effects is spherically symmetric and purely radial. Another assumption is that the core-shell structures can be considered as a homogeneous nanosphere with the radius $R_s$. The equilibrium equation for isotropic and homogeneous materials in absence of body forces provides \cite{17}

\begin{eqnarray} 
G_j\bigtriangledown^2 \mathbf{u}_j + \left(K_j+\frac{G_j}{3} \right)\bigtriangledown(\bigtriangledown \mathbf{u}_j) - \alpha_j K_j\bigtriangledown T = 0, %\rho \ddot{\mathbf{u}},
\label{eq:6}
\end{eqnarray}
where $\alpha$ is the coefficient of thermal expansion, $\mathbf{u} \equiv \mathbf{u}(r)$ is the strain field, $K_j=E_j/3(1-2\nu_j)$ and $G_j=E_j/2(1+\nu_j)$ are the bulk and shear modulus of medium $j$ ($j=c,s,m$), respectively, $\nu_j = (3K_j-2G_j)/2(3K_j+G_j)$ is the Poisson's ratio, and $E_j$ is the Young's modulus. Our calculations use parameters $\nu_c = 0.37$, $K_c = 84$ GPa, $G_{c}=24$ GPa \cite{26}, $\nu_{Fe_3O_4} = 0.37$, $K_{Fe_3O_4} = 227$ GPa, and $G_{Fe_3O_4} = 63$ GPa \cite{27}. Since the shell is a medium and $Fe_3O_4$ (roundly $f$ = 0.4) mixure, the bulk and shear modulus are calculated by $K_s = fK_{Fe_3O_4} + (1-f)K_m \approx 112.8$ GPa and $G_s = fG_{Fe_3O_4} + (1-f)G_m \approx 43.9$ GPa, respectively. The thermal expansion of the shell is expressed by $\alpha_s =f\alpha_{Fe_3O_4}+(1-f)\alpha_m$. The equilibrium equation associated with $\bigtriangledown \times \mathbf{u}_j = 0$ (no rotation) can reduce to be \cite{17}
\begin{eqnarray} 
\bigtriangledown^2 \mathbf{u}_j = \alpha_j\frac{1+\nu_j}{3(1-\nu_j)}\bigtriangledown T(r). %+ \frac{\rho(1-2\nu)(1+\nu)}{(1-\nu)E}\ddot{\mathbf{u}}.
\label{eq:7}
\end{eqnarray}

The deformation field within core $u_c(r)$ and shell $u_s(r)$, and outside ($u_m(r)$) nanoflowers can be obtained by solving Eq.(\ref{eq:7}), using the finiteness of strain field fields at $r = 0$ and $r=\infty$ 
\begin{eqnarray} 
u_c(r) &=& A_cr, \nonumber\\ 
u_s(r) &=& -\alpha_s\frac{1+\nu_s}{3(1-\nu_s)}\frac{\kappa_mT_s}{2\kappa_sR_s^2}\frac{r^3}{5} + A_sr+\frac{B_s}{r^2}, \nonumber\\
u_m(r) &=& \frac{T_sR_s}{2}\frac{\alpha_m(1+\nu_m)}{3(1-\nu_m)}+\frac{B_m}{r^2},
\label{eq:8}
\end{eqnarray}
where coefficients $A_c$, $A_s$, $B_s$ and $B_m$ can found by applying continuity of traction and fields across interfaces. The results imply that the strain displacement outside the nanocomposites decays at infinity and local temperature recovers to room temperature. However, the deformation field within the silver core increases linearly with distance from the origin. Our analysis is in agreement with the previous experimental studies \cite{21}. 

Substituting the strain displacements into the stress-strain relations gives the radial stresses in regions
\begin{eqnarray} 
\sigma_{c,rr}(r) &=& 3K_cA_c, \nonumber\\ 
\sigma_{s,rr}(r) &=& -K_s\alpha_s\frac{3-\nu_s}{1-\nu_s}\frac{\kappa_mT_s}{2\kappa_sR_s^2}\frac{r^2}{5} + 3K_sA_s-\frac{4G_sB_s}{r^3}, \nonumber\\
\sigma_{m,rr}(r) &=& \alpha_m\frac{\nu_mK_m}{1-\nu_m}\frac{T_sR_s}{r} -\frac{4G_mB_m}{r^3}.
\label{eq:10}
\end{eqnarray}
Eq.(\ref{eq:10}) presents the spatial distribution and strong size dependence of the stress strains. While the radial pressure inside the core remain unchanged, the stresses within the nanoflower shell and in medium are strongly dependent on $r$. Under laser irradiation, nanocomposites with larger diameters yield greater light-to-heat conversion than small counterparts \cite{8}. The temperature rise in Eq.(\ref{eq:5}) is currently based on the photothermal effects. Previous studies \cite{10} found that significant heating can be obtained by coupling photothermal approach with hyperthermia effects using AC external magnetic fields. Consequently, the stress strains can be substantially strengthened using both external fields and laser light exposure. The strain fields can be measured by neutron powder diffraction techniques \cite{20}, geometric phase analysis \cite{21,23}, and digital correlation analysis of scanning electron microscope (SEM) images \cite{22}. 

\begin{figure}[htp]
\includegraphics[width=8cm]{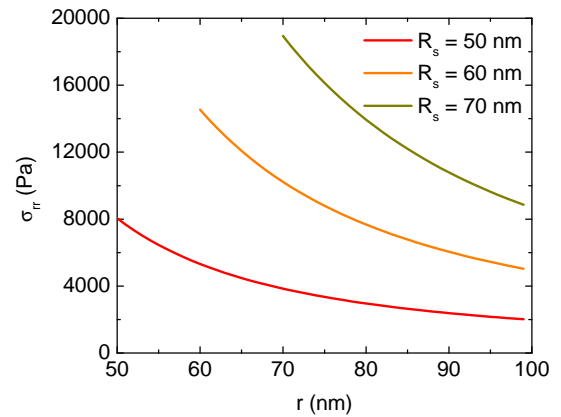}
\caption{\label{fig:3}(Color online) The radial stress outside versus distance from center of Ag@\ce{Fe_3O_4} nanoflowers with various diameters.}
\end{figure}

For water medium, Young's modulus $E_m$ is supposed to be zero, we enable to measure the spatial displacement both stress strains outside nanoflowers. Because $G_m = 0$, $\nu_m = 1/2$, and $T_s$ varies less than $100$ $^0C$, $\sigma_{m,rr}(r) = \alpha_mK_mT_sR_s/r$ is relatively small in comparison with $K_m$. The radial stress is long-ranged and simply inversely proportional to the distance $r$. The manner definitely changes when the surrounding medium is supercooled liquids or solids. Laser-induced thermal stresses on nanoparticle removal from thin films \cite{24} and due to metal nanoparticles in silica matrices \cite{25} are compelling evidence.

When the surrounding medium is silica glass, $\nu_m = 0.17$, $E_m = 73.1$ GPa, $\kappa_m=1.38$ $Wm^{-1}K^{-1}$, and $\alpha_m=0.55\times 10^{-6}$ $K^{-1}$ \cite{16}, we can determine the stress in silica using 400 nm laser irradiation at a power density of $10^4$ $W/cm^2$ for different sizes of the nanocomposites in Fig.\ref{fig:3}. The stress components in silica decay almost as the inverse cube of the distance near the outer surface but it is proportional to $1/r$ in long-range distances. The variation is identical to the previous finding in Ref.\cite{25}. If the stress does not obey an inverse cube law, the system may have impurity, obstacles or asymmetrical interrupting factors. As a result, laser-induced thermoelastic effects can be exploited to detect defects in substances and devices.
\section{Conclusions}
In conclusion, we have explored optical properties and the thermal-induced stress in the surroundings of heated Ag@\ce{Fe_3O_4} nanoflowers in different media using the Mie theory and continuum mechanics theory. Our calculations show that the resonance wavelength of around 400 nm in the absorption is attributed to the surface plasmon resonance of silver and the electronic transition of iron oxide. The second optical band at 620 nm is due to geometrical effects. A subtle interplay between core and shell is supposed to have an inconsiderable effect on the absorption spectrum. The optical peak shift is strongly dependent on the finite size of the nanocomposite and how magnetite is grown on the silver surface. Using laser irradiation or AC magnetic fields leads to temperature rise that generates the strain field inside nanoflowers and their ambient surrounding environment. The thermal stress variation has been analytically found. The long-range stress decays as the inverse of the distance and this finding is in a good agreement with previous study \cite{25}.

\begin{acknowledgment}

%\acknowledgment

This work was supported by Vietnam National Foundation for Science and Technology Development (NAFOSTED) under grant number 103.01-2015.42.

\end{acknowledgment}

\end{document}